# Staircase Selective Harmonic Elimination in Multilevel Inverters to Achieve Wide Output Voltage Range

Amirhossein Pourdadashnia
dept. Electrical and Computer Engineering
Urmia university
Urmia, Iran
st_a.pourdadashnia@urmia.ac.ir

Mohammad Farhadi-Kangarlu
dept. Electrical and Computer Engineering
Urmia university
Urmia, Iran
m.farhadi@urmia.ac.ir

Milad Sadoughi
dept. Electrical and Computer Engineering
Urmia university
Urmia, Iran
st_m.sadughi@urmia.ac.ir

*Abstract*—Multilevel inverters (MLIs) are popular because of their advantages such as improved output voltage quality, lower switching losses, low EMI, and ability to handle higher voltage and power levels. To generate the desired output voltage in the MLIs, there are different switching methods including multicarrier pulse-width modulation (PWM), selective harmonic elimination (SHE), and multilevel space-vector PWM (SVPWM) methods. The SHE method minimizes the number of switching and hence the switching losses. Also, the method eliminates or attenuates specified low-order harmonics and adjusts the fundamental output voltage to the desired value. However, problems arise in the low modulation indexes where the harmonic distortion of the output voltage increases considerably. To solve the problem, this paper proposes a method to adjust the DC voltage when the low output voltage is required. Therefore, the modulation index will increase and the output voltage quality will be improved. The proposed method is verified on a 7-level Cascaded H-Bridge (CHB) inverter using the SHE method solved by particle swarm optimization (PSO) algorithm.

## I. Introduction

Multilevel inverters (MLIs) include an array of dc voltage sources and power semiconductors, which produce stepped waveforms in their output terminal. With the growth of MLIs and their countless advantages, they can be used to connect to high-powered electrical devices. The main advantages of MLIs are high power and voltage, more electromagnetic compatibility, less switching loss, higher efficiency, higher voltage capability, less total harmonic distortion (THD) [1-3]. One of the main challenges in MLI is to achieve quality output voltage as well as eliminate low-order harmonics from the output voltage. Optimization of MLIs in medium and high-voltage industrial applications, such as renewable energy systems, electric motor drives, and flexible transmission equipment (FACTS) has been the subject of much researches [5]. Different topologies are considered for MLIs. The proposed topologies for MLIs include 1) Flying capacitor, 2) Diode clamped inverter, and 3) Cascaded H Bridge (CHB) inverter [4]. The main task of the CHB inverter is to generate and synthesize a waveform, in which the desired output voltage is connected using dc sources, where each output voltage level represents each dc source. The CHB MLI consists of several bridge inverters that are connected in series. Staircase output voltage waveform is the result of connecting the bridges in the MLIs inverter.

Several methods for pulse width modulation (PWM) have been proposed to achieve the performance of output voltage control and simultaneous reduction of unwanted harmonics, including sinusoidal PWM carrier techniques (SPWM) and spatial vector modulation (SVM) techniques. However, conventional techniques such as PWM cannot eliminate low-level harmonics effectively. Selective harmonic elimination (SHE) is a practical and effective way to eliminate or attenuate low-order harmonics. SHE technique can cause the inverter output voltage to be of good quality and lower switching frequency by effectively solving the SHE equations compared to other modulation methods. Solving the SHE equations is more challenging. The SHE method involves nonlinear equations, in other words, the equations contain trigonometric expressions that make it more difficult to solve the SHE equations [5]. The particle swarm optimization (PSO) is a powerful algorithm for solving the required parameters of the SHE technique equations. When the selected annoying harmonics are removed, we get the expected voltage. At the desired voltage, the individually selected harmonics are eliminated or minimized. In the proposed idea, to achieve the desired voltage range, in the modulation index less than 0.5 per unit, the values of voltage sources are reduced by half. [6]. In the proposed method, at low modulation coefficients, the amount of low-order harmonics is reduced or eliminated. The quality of the voltage waveform generated by the CHB inverter as well as the THD is significantly improved. Using this method, we achieve wide voltage ranges with high-quality. To operate the inverter at low modulation indices using the proposed method, the voltage stress on the switches and also filter size are considerably reduced [8].

In this paper, the SHE equations of the proposed method are solved using the PSO algorithm and the switching angles for a 7-level inverter are calculated. The proposed method provides a wide range of voltages with low THD.

## II. Proposed Method for SHE in CHB Inverters

### A. Multilevel Inverters

With the advent of the third level in electronic converters, the term multilevel becomes more meaningful. The main concept of the MLI is the use of several dc sources to carry out power conversion through the synthesis of the phase voltage waveform, with the feature of achieving higher power [8], [9]. Special features of CHB MLIs such as the simplicity of the control method, use of independent dc sources, and modularity have caused more attention and application. They are also widely used for the following reasons:

1. The minimum number of components required compared to a diode clamped and a flying capacitor to achieve the same voltage level.

2. There are no additional clamped diodes or voltage balancing capacitors.

3. Soft switching methods can be used to reduce both switching losses and dv/dt.

4. Having high modularity, which means in case of breakdown, reduces the repair time as well as maintains continuity of the power supply.

5. As the output levels increase, the harmonic distortion decreases. The disturbing harmonics of the output voltage waveform decrease with an increasing modulation index. Therefore, the output waveform significantly became a sinusoidal waveform, and as a result, the harmonic distortion is reduced.

The typical structure of a single-phase CHB inverter is shown in Fig. 1.

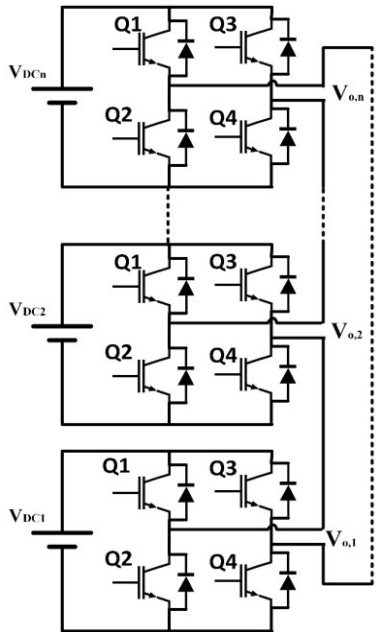

Fig. 1.Typical structure of single phase CHB inverter.

### B. Mathematical Model of SHE

The frequency of each harmonic component is a multiple of its principal component, and these components can be extracted using the Fourier transform.

Assuming the symmetry of dc sources and the amplitude of equality of all dc sources, to obtain the formula, the Fourier series analysis of the stair output voltage waveform is given by (1).

$$v(\omega t) = v_n(\alpha) \sin(n\omega t) \qquad (1)$$

where

$$v_{n(\alpha)} = \frac{4V_{dc}}{n\pi} \sum_{k=1}^{s} \cos(n\alpha_k) \qquad (2)$$

For odd n
And

$$v_n(\alpha) = 0 \text{ for even n} \qquad (3)$$

Tripled harmonics is one of the problems of single-phase power systems. To solve the problem of the existence of the tripled harmonic, in the equations of the technique of removing the selected harmonics, we consider the third harmonic. By combining (1), (2), and (3):

$$v(t) = \sum_{n=1,3,5,\ldots}^{\infty} \left\{ \frac{4v_{dc}}{n\pi} \left[ \cos(n\alpha_1) + \cos(n\alpha_2) + \ldots + \cos(n\alpha_s) \right] \sin(n\omega t) \right\} \qquad (4)$$

Subject to $0 < \alpha_1 < \alpha_2 < \ldots < \alpha_s < \frac{\pi}{2}$

also, S is the number of switching angles and n is the harmonic order.

$$v(\omega t) = v_1 \sin(\omega t) \qquad (5)$$

Also, the fundamental voltage $V_1$ can be written as:

$$v_1 = \frac{4V_{dc}}{\pi}(\cos(\alpha_1) + \cos(\alpha_2) + \cos(\alpha_3)) \qquad (6)$$

The modulation index (M) can be defined and evaluated according to Equation (8). For a simpler description of the proposed method and the classical method, instead of the M, we use the parameter per unit value of voltage.

When all the switching angles according to equations (4) to (6) are zero, the maximum output voltage of the inverter is obtained.

$$v_{1\max} = \frac{4SV_{dc}}{\pi} \qquad (7)$$

$$m_i = \frac{v_1}{v_{1\max}} = \frac{\pi v_1}{4SV_{dc}} \qquad (8)$$

$$v_1 = m_i \left(\frac{4SV_{dc}}{\pi}\right) for\, 0 < m_i \leq 1 \qquad (9)$$

Also, by solving the following nonlinear equations known as SHE equations, the switching angles that lead to the synthesis of the output voltage of the CHB inverter with a minimum THD are obtained:

$$\frac{4V_{dc}}{\pi}(\cos(\alpha_1) + \cos(\alpha_2) + \cos(\alpha_3)) = v_1$$
$$\cos(3\alpha_1) + \cos(3\alpha_2) + \cos(3\alpha_3) = v_3 \qquad (10)$$
$$\cos(5\alpha_1) + \cos(5\alpha_2) + \cos(5\alpha_3) = v_5$$

$v_3$ and $v_5$ represent harmonic 3rd and 5th, respectively, which are derived from (10) [6].

Fig. 2 illustrates the output waveform of the CHB inverter.

The output voltage waveform of each CHB inverter is generated by creating a duty cycle at the ac terminal. Fig. 2 also shows the output voltage of the inverter, in which the number of steps is 2S + 1, and D is the number of isolated dc sources on an inverter-based of the CHB [8].

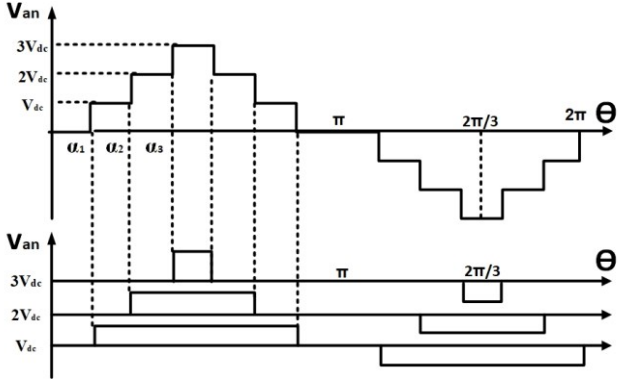

Fig. 2. staircase output voltage waveform of a single-phase CHB inverter.

## C. Particle Swarm Optimization Algorithm

PSO is an efficient and modern optimization algorithm developed by two researchers named Eberhart and Kennedy in 1995 [10]. In the PSO algorithm, each particle of the population (particles) moves in a simulated space for the problem, which can be multidimensional. Then, the algorithm finds the desired point and move the particle to its new position. Eventually, reaches the best location for the particle by updating the position of the particle according to experience. In the PSO algorithm, population parameters are randomly determined. The velocity of a particle is affected by three parameters, which include the components of inertial motion components, social components, and cognitive components. The inertia parameter models the inertial behavior of particles to move in the previous direction. The social component refers to the memory of particles from the best position among particles. Each factor defines its best value so far with the parameter ($p_{best}$), and also the best value in the group with the parameter ($g_{best}$) between the particles of the algorithm.

By mathematical definitions, the velocity update equation is as follows:

$$v_{i+1} = \omega \times v_i + c_1 \times r_1 \times (P_{best_i} - x_i) + c_2 \times r_2 \times (g_{best_i} - x_i) \tag{11}$$

By analysing (11), we find that the parameter $V_i$ is the current velocity of the particle, the $X_i$ the parameter is the current position of the particle, the parameter $\omega$ is the weight of inertia, $C_1$ is the personal acceleration index, $C_2$ is the social acceleration factor, the equation is the best particle position, the equation is the best global position Among the group is particles and $r_1$, $r_2$ are accidental numbers.
The position update equation is given as follows:

$$x_{i+1} = x_i + x_{i+1} \tag{12}$$

Fig. 3 shows the iteration scheme of the particles for the PSO algorithm. Fig. 3 demonstrates the step-by-step steps of the algorithm execution schematically.

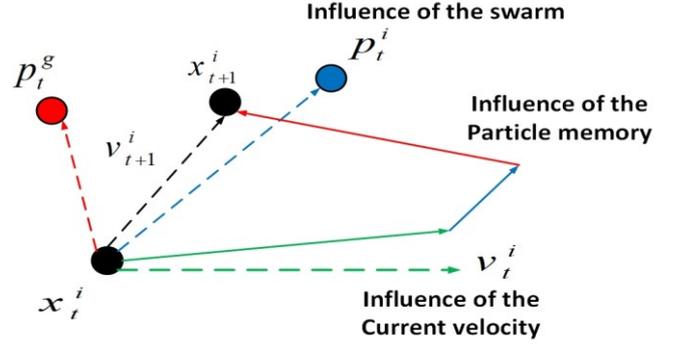

Fig. 3. Iteration scheme of the particles.

Fig. 4 shows the angles obtained by solving the SHE equations with the PSO algorithm for a 7-level multilevel inverter for each output voltage level in the classical method. Fig. 5 also shows the new angles obtained for the proposed method. In Fig. 5, for the output voltage index per unit, less than 0.5 angles are calculated with the proposed method.

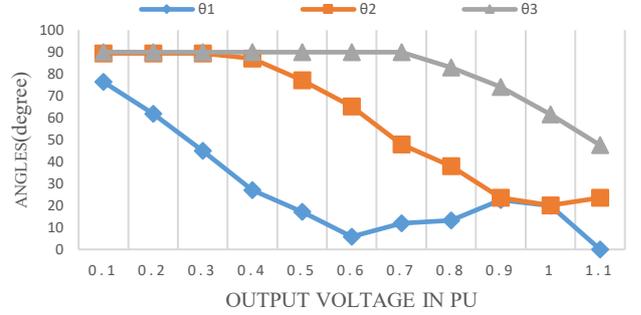

Fig. 4 output voltage in pu versus switching angles in the classic method.

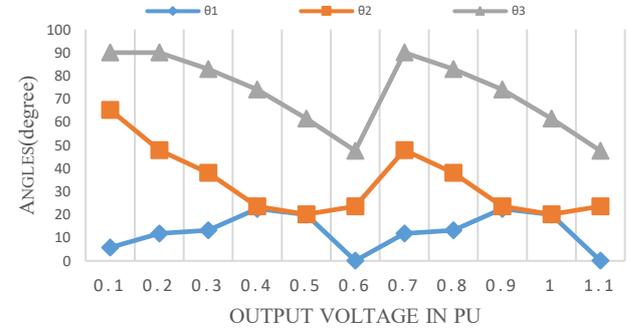

Fig. 5. output voltage in pu versus switching angles in the proposed method.

### III. SIMULATION RESULT

#### A. The Classic Method of SHE

In this section, we define the cost function for the SHE-PWM technique. The cost function equations are expressed as follows:

$$f = A \times \left| M - \frac{|v_1|}{DV_{dc}} \right| + B \times \sum_{i=2}^{S} \frac{1}{h_{si}} \frac{|v_{si}|}{DV_{dc}} \tag{13}$$
$$i = 2, 3, ..., S$$

$$M = \frac{|v_1|}{s v_{dc}} \quad (14)$$

Parameters A and B describe the weight of the basic component and selected harmonics, severally, and also M demonstrate the modulation index component of the objective function [9]. For a simpler description of the proposed method and the classical method, instead of the modulation index, we use the parameter per unit value of voltage. D is the number of dc sources, and the parameter S is the number of switching angles. $v_{si}$ is the number of undesirable harmonics and $\frac{1}{h_{si}}$ is the unwanted harmonic sequence in the inverter output voltage. Actually, consider more weight for low-order harmonics and less weight for high-order harmonics. Using this proposed cost function, the basic harmonic amplitude of the output voltage of the CHB inverter is calculated correctly and the selected annoying harmonics are eliminated. The main purpose is to solve the SHE nonlinear equations, which are set to obtain the optimal switching angles. The optimal switching angles for the MLI were selected to produce a stepped voltage at inverter output with lower THD.

First, the formulas of the SHE technique are applied to the PSO algorithm, then the angles are calculated. In the Simulink MATLAB, a 7-level CHB inverter is simulated, then the angles obtained from the PSO algorithm are applied in the Simulink MATLAB simulated inverter software. A wide range of voltages is obtained where the low-order harmonics in the output voltage waveform are eliminated or reduced. The voltage diagram in terms of output voltage per-unit is shown in Fig. 6. In which the expected voltage is equal to the voltage obtained by applying the switching angles obtained from the PSO algorithm.

Fig. 6 shows the Voltage diagram in terms of $V_{o, pu}$, which its base value in this paper is 300 volts.

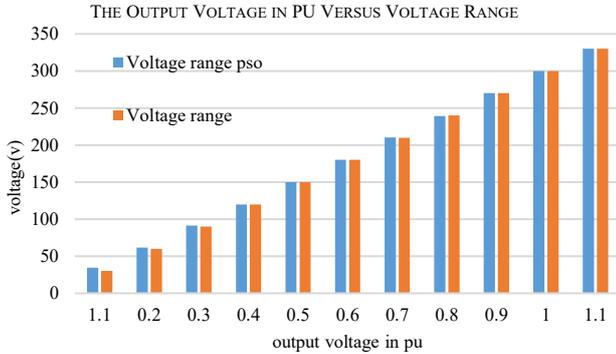

Fig. 6. Voltage diagram in terms of output voltage in pu.

The The voltage diagram clearly shows that the voltage range obtained corresponds exactly to the expected voltage.

Fig. 7 and Fig. 9 show the output voltage waveforms in two per unit values of voltage coefficients of 0.1 and 0.5, respectively.

According to Fig. 8 and Fig. 10, by analysing the output voltage waveform (FFT), it can be seen that in the value of voltage ranges less than 0.5, the magnitude of low-order harmonics is high. In the other words, their THDs are extremely high.

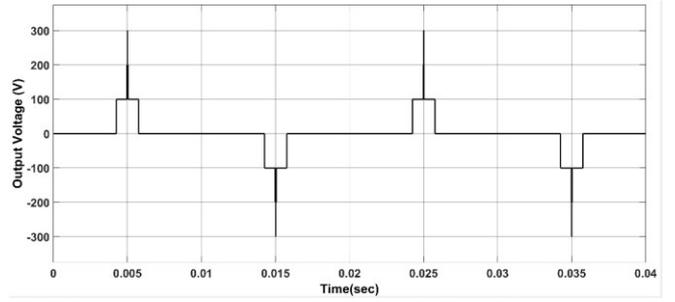

Fig. 7. Inverter phase output voltage waveform in $V_{o, pu}$ (0.1).

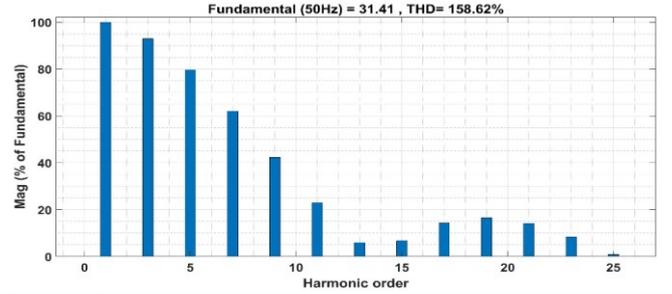

Fig. 8. FFT analysis of inverter phase voltage waveform in $V_{o, pu}$ (0.1).

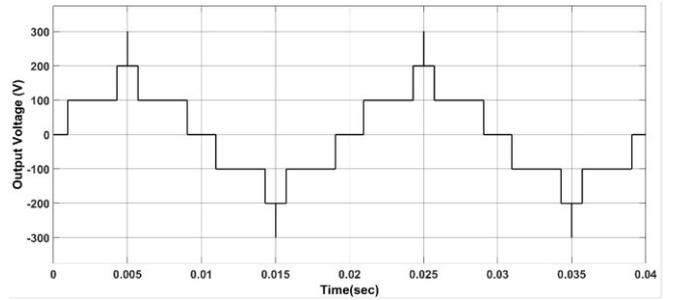

Fig. 9. Inverter phase output voltage waveform in $V_{o, pu}$ (0.5).

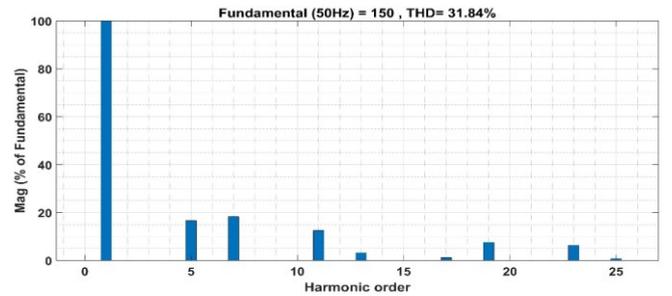

Fig. 10. FFT analysis of inverter phase voltage waveform in $V_{o, pu}$ (0.5).

By the fact that increasing the modulation index results that more low-order harmonics can be eliminated is clearly shown in Fig. 10. In Figs. 10 and 12, the actual voltage amplitude is also obtained correctly. As expected, we find that when the per-unit value of voltage rise, the output voltage turns into a sinusoidal waveform. The efficiency of the SHE method is based on PSO algorithm and provides different voltage levels with a low THD, but the main problem of the classic method is the existence of low-order harmonics at the modulation index below $V_{o, pu}$ 0.5. In the next section, an innovative method is proposed to solve this problem.

## B. The Proposed Method for Eliminating Selective Harmonics

Analysing the results obtained in the previous section, it was observed that at the low $V_{o,pu}$ especially less than 0.5, the low-order harmonics are not completely removed, which causes be an increase in the THD. To dissolve this subject, we use the proposed method by halving the dc sources. The proposed method causes the voltage range to be obtained correctly and the low-order harmonics to be significantly reduced. In this method, to reach the expected voltage range in the classical case, twice the modulation index must be used to calculate the switching angles, which is also due to the halving of the voltage sources. This proposed method, due to the reduction of input voltage, also has the advantage of reducing the voltage stress on inverter elements such as switches, which reduces the complexity and enhanced economic benefits.

Equation (17) describes the proposed method, in which the modulation index has doubled due to the halving of the dc-link voltage.

$$V_{o,pu} = S \times V_{dc} \times M_{old} \tag{15}$$

$$V_{o,pu} = S \times \frac{V_{dc}}{2} \times M_{new} \tag{16}$$

$$S \times V_{dc} \times M_{old} = S \times \frac{V_{dc}}{2} \times M_{new} \tag{17}$$

$$M_{new} = 2 \times M_{old}$$

In the following, the waveform analysis (FFT) of the voltage waveform obtained in the $V_{o,pu}$, 0.1 and 0.5 with the method proposed in this paper is shown.

According to Fig. 12 and Fig. 14 by analysing the output voltage waveform (FFT) voltage waveform and comparing them with the classical method, it is seen that in the proposed method, the two selected 3rd and 5th harmonics are significantly weak.

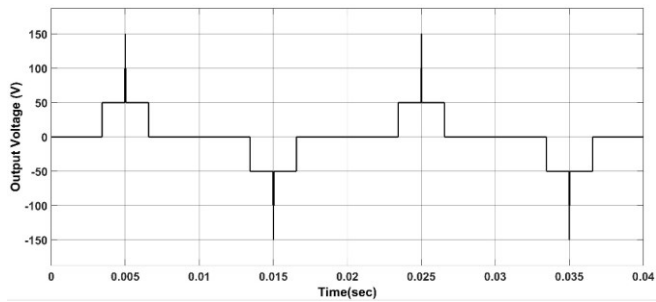

Fig. 11. Inverter phase output voltage waveform in $V_{o,pu}$ (0.1).

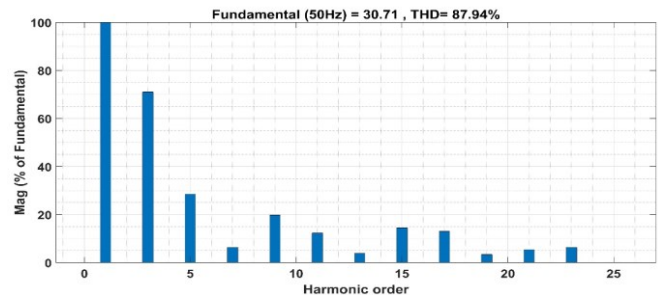

Fig. 12. FFT analysis of inverter phase voltage waveform $V_{o,pu}$ (0.1).

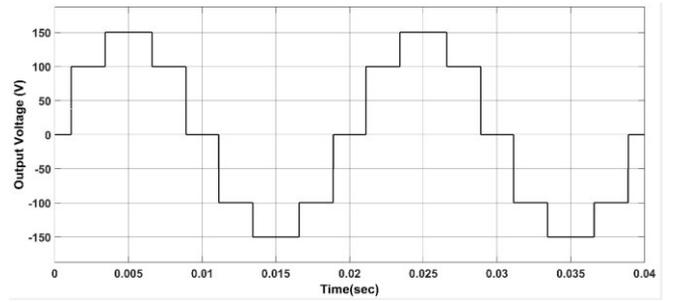

Fig. 13. Inverter phase output voltage waveform in $V_{o,pu}$, (0.5).

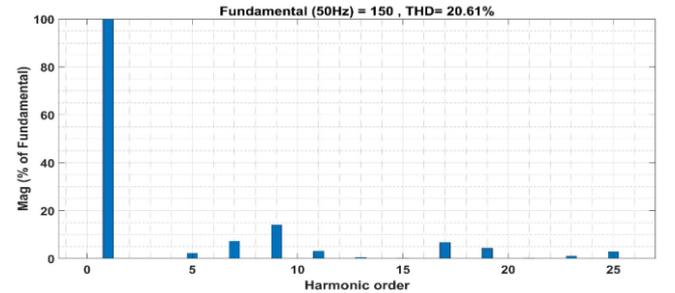

Fig. 14. FFT analysis of inverter phase voltage waveform in $V_{o,pu}$, (0.5).

According to Fig. 15, due to halving the input voltage sources, twice the $V_{o,pu}$, is used compared to the classical method. The proposed method improves the index (THD) in the voltage coefficient of $V_{o,pu}$, lower than 0.5 compared to the classical method

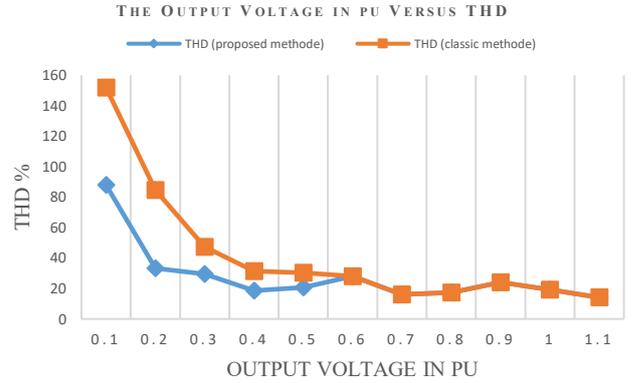

Fig. 15. THD% diagram versus per unit value of voltage.

The orange diagram shows the harmonics of the classical method and the blue diagram show the harmonics of the proposed method.

In Table .1 for $V_{o,pu}$, below 0.5, the value of the obtained voltage waveform index (THD) is compared with the percentage of improvement of this index expressed as a percentage. This table clearly shows that the original idea of the paper is correct.

Table 1. Comparison of THD% output voltage waveform in the studied method.

| s.no | Per unit value of voltage | THD% Classic Method | THD% proposed Method | The rate of improvement |
|------|---------------------------|---------------------|----------------------|-------------------------|
| 1 | 0.1 | 158.62 | 87.94 | 43.92% |
| 2 | 0.2 | 84.66 | 33.23 | 60.74% |
| 3 | 0.3 | 47.29 | 29.59 | 37.43% |
| 4 | 0.4 | 31.29 | 18.66 | 40.37% |
| 5 | 0.5 | 30.41 | 20.61 | 32.88% |

## IV. Conclusion

In this paper, a SHE-PWM method for MLIs is introduced to increase the output voltage quality for low output voltage values. This method is based on reducing the DC voltage at low output voltage performance. The efficiency of the proposed method was confirmed by simulation studies. As the results showed, the harmonic distortion of the output voltage in the proposed method is significantly reduced. For example, THD of output voltage for the classical method was 158.62% and the proposed method was 94.87% at the output voltage of 30 volts. This method greatly improves the harmonic distortion to achieve low voltage amplitudes.

## V. References

[1] E. Babaei, M. Farhadi Kangarlu, and F. Najaty Mazgar, "Symmetric and Asymmetric Multilevel Inverter Topologies with Reduced Switching Devices," Elsevier Journal of Electric Power Systems Research, vol. 86, pp. 122-130, May 2012.

[2] L. M. Tolbert, F. Z. Peng, and T. G. Habetler, "Multilevel converter for large electric drives," IEEE Trans. Ind. Appl., vol. 35, no. 1, pp. 36–44, Jan./Feb. 1999.

[3] D. Parrott and X. Li, "Locating and tracking multiple dynamic optima by a particle swarm model using speciation," IEEE Transactions on Evolutionary Computation, vol. 10, no. 4, pp. 440‐458, Jul. 2006.

[4] M. Farhadi Kangarlu and E. Babaei, "A Generalized Cascaded Multilevel Inverter Using Series Connection of Submultilevel Inverters," in IEEE Transactions on Power Electronics, vol. 28, no. 2, pp. 625-636, Feb. 2013, doi: 10.1109/TPEL.2012.2203339.

[5] M. T. Hagh, H. Taghizadeh, and K. Razi, "Harmonic minimization in multilevel inverters using modified species-based particle swarm optimization," IEEE Transactions on Power Electronics, vol. 24, no. 10, pp. 2259‐2267, Oct. 2009.

[6] M. Farhadi Kangarlu and E. Babaei, "Variable DC voltage as a solution to improve output voltage quality in multilevel converters," 4th Annual International Power Electronics, Drive Systems and Technologies Conference, Tehran, 2013, pp. 242-247, doi: 10.1109/PEDSTC.2013.6506711.

[7] H. Taghizadeh and M. Tarafdar Hagh, "Harmonic elimination of multilevel inverters using particle swarm optimization," 2008 IEEE International Symposium on Industrial Electronics, Cambridge, 2008, pp. 393-396, doi: 10.1109/ISIE.2008.4677093.

[8] E. Babaei, M. Farhadi Kangarlu, and M. Sabahi, "Mitigation of voltage disturbances using dynamic voltage restorer based on direct converters," IEEE Trans. Power Del., vol. 25, no. 4, pp. 2676-2683, Oct. 2010

[9] E. Babaei, M. F. Kangarlu, and M. A. Hosseinzadeh, "Asymmetrical multilevel converter topology with reduced number of components," IET Power Electronics, vol. 6, no. 6, pp. 1188‐1196, Aug. 2013.

[10] J. Kennedy, R.C. Eberhart, "Particle Swarm Optimization," Proc. IEEE Int. of. Neural Networks, Piscataway, NJ, USA, 1942-1948, 1995.

[11] E. Babaei, M. F. Kangarlu, M. Sabahi, and M. R. A. Pahlavani, "Cascaded multilevel inverter using sub-multilevel cells," Electric Power Systems Research, vol. 96, pp. 101‐110, 2013.

[12] E. Babaei, M. F. Kangarlu, and M. Sabahi, "Dynamic voltage restorer based on multilevel inverter with adjustable dc-link voltage," IET power electronics, vol. 7, no. 3, pp. 576‐590, Apr. 2013.